\documentclass[a4paper,11pt]{article}
\usepackage{amsmath,amssymb}
\usepackage{graphicx}
\def\be{\begin{equation}}
\def\ee{\end{equation}}

\def\bea{\begin{eqnarray}}

\def\eea{\end{eqnarray}}

\def\ben{\begin{displaymath}}
\def\een{\end{displaymath}}
\def\ba{\begin{array}{c}}
\def\bal{\begin{array}{l}}
\def\ea{\end{array}}

\begin{document}

%.

\vspace{1.5cm}

 \begin{center}{\Large \bf
Thermodynamics of Pseudo-Hermitian Systems in Equilibrium
  }\end{center}

\vspace{10mm}

 \begin{center}

 {\bf V\'{\i}t Jakubsk\'{y}}

 \vspace{3mm}

 \'{U}stav jadern\'e fyziky AV \v{C}R

250 68 \v{R}e\v{z}

Czech Republic

{e-mail: jakub@ujf.cas.cz}
\end{center}
\begin{abstract}
In study of pseudo(quasi)-hermitian operators, the key role is played by the positive-definite metric operator. It enables physical interpretation of the considered systems. In the article, we study the pseudo-hermitian systems with constant number of particles in equilibrium. We show that the explicit knowledge of the metric operator is not essential for study of thermodynamic properties of the system. We introduce a simple example where the physically relevant quantities are derived without explicit calculation of either metric operator or  spectrum of the Hamiltonian.
\end{abstract}

\section{Alternative representation of hermitian operators}

In quantum mechanics, the central role is played by unitary transformations. They leave scalar product and norm of states unchanged.  Particularly, Hamiltonians are required to be hermitian in order to keep unitarity of time evolution. In this section, we want to examine how the theory should be modified to keep the norm conserved when a wider class of transformations is considered. 
\subsection{Non-trivial scalar products}
Let us consider a hermitian Hamiltonian $H$ with discrete spectrum which acts on ${\cal{H}}=({\cal V}, (.,.))$. Let the operator ${B}$ be a non-unitary bijection of ${\cal H}$ to itself (and bijection on ${\cal V}$ particularly)
 \begin{equation} {\cal B}^{\dagger}\neq {\cal B}^{-1}.\label{Bnonun}\end{equation}
Let us use $\cal B$ as a similarity transformation 
 \begin{equation}\tilde{H}={\cal B}^{-1}H{\cal B},\ \ \tilde{\psi}={\cal B}^{-1}\psi\in\tilde{\cal H},\ \ \psi\in{\cal H}\label{Btrans}. \end{equation}
It leaves the vector space $\cal V$ unchanged. However,
the Hamiltonian  $\tilde H$
ceases to be hermitian on $\cal H$. Instead, it satisfies the following relation 
 \begin{equation} \tilde{H}^{\dagger}=\Theta\tilde{H}\Theta^{-1} ,\label{quasi}\end{equation}
where the operator $\Theta$ is hermitian, strictly positive and has of the following explicit form
 \begin{equation}\Theta={\cal B}^{\dagger}{\cal B},\ \ \Theta=\Theta^{\dagger},\ \ \ \Theta>0.\label{Theta}\end{equation}
\noindent
The norm of states is not preserved under the similarity transformation (\ref{Btrans}). There holds the following relation instead
 \begin{equation}||\psi||^2=(\psi,\psi)=({\cal B}\tilde{\psi},{\cal B}\tilde{\psi})=(\tilde{\psi},\Theta\tilde{\psi}).\label{normnotconv}\end{equation}
\noindent
To keep the norm conserved, we define an alternative scalar product $(.,.)_{\Theta}\equiv(\ .\ ,\Theta\ .\ )$ on the vector space ${\cal V}$. 
Consequently, the operator $\cal B$ may be considered as a unitary mapping between two Hilbert spaces $\cal H$ and $\tilde{\cal H}$ which coincide in vector space $\cal V$ but Hilbert space structure of $\tilde{\cal H}$ is induced by $(.,.)_{\Theta}$. Particularly, $\tilde H$ is hermitian in $\tilde{\cal H}$. We may rewrite (\ref{normnotconv}) as
\begin{equation} ||\psi||^2=|| \tilde{\psi}||^2_{\Theta}.\end{equation}
\noindent

To avoid possible misunderstanding caused by the alternative definition of scalar product, let us call the representation $({H},{\cal H})$ as a hermitian picture while we will call the representation $(\tilde{H},\tilde{\cal H})$ as a quasi-hermitian picture. The description of a physical system in both representations is equivalent as these are related by the unitary mapping  ${\cal B: H}\mapsto\tilde{{\cal H}}$.
 
The advantage of quasi-hermitian picture lies in extended freedom in choice of $\cal B$. It may be exploited to simplify the Hamiltonian and determine its spectral properties more easily. On the other hand, we lose an intuitive interpretation for the operators. For instance, the coordinate operator in the quasi-hermitian picture
 \begin{equation}\tilde{X}={\cal B}^{-1}X{\cal B}\label{x}\end{equation}
may have quite complicated form dependently on the choice of $\cal B$.

Concluding, the effort to employ wider class of transformations led to the construction a new Hilbert space with non-trivial metric operator.
In this context, we may refer to many-body physics where the alternative representation is used in boson realization of fermionic systems \cite{geyerscholzhahne,Geyer}. The non-Hermiticity of resulting bosonic Hamiltonian is treated by redefinition of the scalar product which fits to the strategy outlined above.

\subsection{Statistical description of equilibrium systems }

Let $H$ describes an ensemble of a huge but fixed number of independent particles. Let us focus to the statistical description of the system. The state of the system is characterized by a density matrix $\rho$ which can be written in energy representation as
 \begin{equation}\rho=\sum_n W_n|n\rangle\langle n|,\ \ \mbox{Tr}\rho<\infty.\label{rho}\end{equation}
The coefficients $W_n$ represents (up to normalization) the probability with which we may find the system in the state $|n\rangle$. In what follows, we will consider a system in equilibrium. In that case, the density matrix operator is solution of Bloch equation
 \begin{equation}\frac{\partial \rho}{\partial \beta}=-H\rho\label{bloch}\end{equation}
with initial condition $\rho(0)=1$. 
Its formal solution
 $\rho=e^{-\beta H}\ $ 
fixes the coefficients to be $W_n=\exp(-\beta E_n)$. Normalization factor of the density matrix depends on the inverse temperature $\beta$ and is called partition function
 \begin{equation} Z=Tr\rho.\label{Z}\end{equation}
It plays a crucial role in thermodynamics of the system as it allows direct computation of thermodynamic quantities.

Let us observe how is the description modified in quasi-hermitian picture.
Once we have (\ref{rho}) satisfying (\ref{bloch}), the density operator in quasi-hermitian picture is 
 \begin{equation}\tilde{\rho}={\cal B}^{-1}\rho{\cal B}=\sum_{n}\exp(-\beta E_n)|\tilde{n}\rangle{\langle\langle  \tilde n|},\label{rhotilda}\end{equation}
where 
$$\langle\langle \tilde{n}|\tilde{H}=E_n\langle\langle\tilde{ n}|,\ \ \tilde{H}|\tilde{ n}\rangle=E_n|\tilde{ n}\rangle,\ \ \Theta|\tilde{ n}\rangle=|\tilde {n}\rangle\rangle,\ \ \langle \tilde{n}|\tilde{m}\rangle_{\Theta}=\langle\langle\tilde{n}|\tilde{m}\rangle=\delta_{m,n}$$
and satisfies 
$\partial_{\beta}\tilde{\rho}=-\tilde{H}\tilde{\rho}$ as long as $\cal B$ does not depend on $\beta$. 
Let us derive the partition function $\tilde{Z}$ for (\ref{rhotilda}).
Partition function (\ref{Z}) may be written as
 \begin{equation} Z=\sum_n\langle n|\rho|n\rangle=\sum_n\langle n|{\cal B}\tilde{\rho}{\cal B}^{-1}|n\rangle=\sum_n \langle\tilde{n}|{\cal B}^{\dagger}{\cal B}\tilde{\rho}|\tilde{n}\rangle=\sum_n\langle\tilde{n}|\tilde{\rho}|\tilde{n}\rangle_{\Theta}=\tilde{Z}.\label{Ztilda}\end{equation}
Thus the partition functions in both pictures coincide and do not depend on the mapping $\cal B$. A similar conclusion holds for mean value of observable $A_i$
 \begin{equation}\langle A_i \rangle_{\rho}=\mbox{Tr}\left(A_i\rho\right)=\mbox{Tr}\left(\tilde{A}_i\tilde{\rho}\right)=\langle \tilde{A}_i \rangle_{\tilde{\rho}}\label{mean}\end{equation}
as long as $A_i\rho$ is a trace-class operator.

Bloch equation (\ref{bloch}) resembles Schrodinger equation for evolution operator. The link can be established by the substitution $\beta\rightarrow it$. Then (working in quasi-hermitian picture) the equation (\ref{bloch}) reads
 \begin{equation} it\frac{\partial \tilde{U}}{\partial t}=\tilde{H}\tilde{U},\ \ \tilde{U}(0)=1\label{propagator}\end{equation}
where we defined $\tilde{U}(t)=\tilde{\rho}(it)$. In $x$-representation, we have     \begin{equation}\tilde{\rho}(x_1,x_2)|_{\beta\rightarrow it}=\tilde{U}(x_1,x_2;t)= \langle x_1| e^{-i\tilde{H}t}|x_2 \rangle\equiv: G(x_1,x_2;t).\label{propagator2}\end{equation}
We should be careful with the interpretation of $G$. Although it is manifestly similar to a quantum mechanical propagator, there is a difference we should keep in mind.
%As the Hamiltonian $\tilde{H}$ is non-hermitian with respect to the standard scalar product, the evolution is non-unitary. 
The vectors $|x\rangle$ are not eigenstates of the position operator in the quasi-hermitian picture. This role is played by $|\tilde{x}\rangle={\cal B}^{-1}|x\rangle$, see (\ref{x}). 

Despite the lack of intuitive interpretation, the "propagator" (\ref{propagator2}) is of importance as it allows to compute partition function ($Z=\int G(x,x;-i\beta)dx$) with use of standard techniques. Particularly, let us mention standard perturbation theory or path integrals in this context. Dependently on the technique used, the spectrum of $\tilde{H}$ is not essential for computation of the partition function.

\section{Pseudo-hermitian operators}

\subsection{A bit of history
}

Towards the end of the last century an interest of a few groups of mathematicians 
has been attracted by the rather unexpected behavior of certain  manifestly 
non-hermitian
anharmonic-oscillator 
Hamiltonians
which seemed to generate the purely real spectra \cite{Caliceti, Buslaev&Grecchi}.
In the next stage of development,
it was Bender and Boettcher \cite{Bender&Boettcher1,Bender&Boettcher2}
who revealed the possible deeper relevance
of the similar models 
(well exemplified by the ``zero mass" Hamiltonian
 \be 
 \tilde{H}=p^2+i x^3\label{ix^3}
 \ee
without free parameters)
in the context of physics.
Subsequently, the reality of the spectra of the similar models has 
been tested  \cite{Pham} and, finally, rigorously
proved \cite{DDT}.

As long as one has  $[\tilde{H},PT]=0$ in eq. (\ref{ix^3}),
(where the operators $P$ and $T$ are chosen as representing the standard space reflection and time reversal, respectively),
the authors of
ref. 
\cite{Bender&Boettcher1,Bender&Boettcher2} 
conjectured
that
the reality of the spectra of the 
manifestly non-hermitian Hamiltonians of the form  (\ref{ix^3}) may be attributed to their peculiar 
anti-linear symmetry called  $PT$-symmetry (re-interpreted as 
$P$-pseudohermiticity in ref.  \cite{Mostafazadeh1,Mostafazadeh2}).
This idea initiated an intensive study of  pseudo-hermitian operators with anti-linear symmetries \cite{PTstuff1,PTstuff2,PTstuff3,PTstuff4,PTstuff5} where the original specific
choice of the parity is being extended to any 
invertible hermitian mapping $\eta$ such that the operator equation
 \be \tilde{H}^{\dagger}\eta=\eta \tilde{H}\label{pseudoeta}\ee
is satisfied. In such a case the spectrum of 
the corresponding pseudo-hermitian, diagonalizable operator is either purely real or it contains the complex conjugate pairs \cite{Bender&Boettcher1,Bender&Boettcher2,Mostafazadeh1,Mostafazadeh2}.

The operator $\eta$ is non-unique and represents rather a class of generally indefinite operators satisfying (\ref{pseudoeta}). However, it revealed that when the spectrum of the Hamiltonian is purely real, the class contains positive definite operators and, consequently, a consistent probabilistic interpretation (including unitary time evolution) of state vectors associated with pseudo-hermitian Hamiltonian is enabled \cite{Q,Cop,theta}. With a special denotation of these operators, the equation (\ref{pseudoeta}) reads now 
 \begin{equation} \tilde{H}^{\dagger}\Theta=\Theta \tilde{H},\ \ \Theta>0.\label{htth}\end{equation}

Due to its positive definiteness, the metric operator can be decomposed $\Theta={\cal B}^{\dagger}{\cal B}$. In this point, we may feel inspired by the previous section and say that the operators $\tilde{H}$ and $\Theta$ establish quasi-hermitian picture of a physical system. The hermitian picture $H={\cal B}\tilde{H}{\cal B}^{-1}$ may provide here an intuitive insight in the nature of interaction which ceases to be transparent in the quasi-hermitian picture \footnote{In (\ref{ix^3}), $x$ and $p$ do not represent position and momentum operators. These are  $\tilde{x}={\cal B}^{-1}x{\cal B}$ and $\tilde{p}={\cal B}^{-1}p{\cal B}$ in quasi-hermitian picture.}. Moreover, additional physical observables in quasi-hermitian picture may be defined with help of appropriate operators in hermitian picture, see eq. (\ref{x}).

Let us emphasize, that the current treatment differs from the previous section. The relation between hermitian and quasi-hermitian representation is more complicated as the operator $\Theta$ (and $\cal B$) is not given explicitly in the present case. Its construction depends on the Hamiltonian $\tilde{H}$ and the resulting metric operator is generally non-unique. There exists rather a class of $\Theta$`s which satisfy (\ref{htth}) for given $\tilde{H}$. The construction and the associated ambiguity of the scalar product $(.,.)_{\Theta}$ represent the main questions within the theory and are tackled in various ways, see e.g. \cite{geyerscholzhahne,MostafazadehKBJakSmeBenderQ1,MostafazadehKBJakSmeBenderQ2,MostafazadehKBJakSmeBenderQ3}. 

%We shall compare (\ref{htth}) with (\ref{quasi}).  We may introduce a new scalar product $(.,.)_{\Theta}$ which constitute together with $\tilde{H}$ a quasi-hermitian picture of a physical system. In the spirit of the previous section, we may conclude that non-hermitian operator with real spectrum define a hermitian counterpart implicitly.%which would be represented by a hermitian Hamiltonian $H$ in the standard Hilbert space. 

The conclusions of the previous section find their important application here. We showed (\ref{Ztilda}) that the partition function is the same in both pictures. Thus we may compute $Z$ in quasi-hermitian picture where the Hamiltonian is given explicitly. This enables direct computation of thermodynamic properties of a physical system in equilibrium described by quasi-hermitian without explicit knowledge of the metric operator $\Theta$. 

In the next section, we intend to present a simple example of quasi-hermitian system. We derive its thermodynamic characteristics independently on both the metric operator and the spectrum of the Hamiltonian.

\subsection{Thermodynamics of $2\times 2$ toy model}

Let us consider an ensemble of $N$ distinguishable particles neglecting their mutual interaction. The partition function may be factorized
 \begin{equation} Z =\left(Z_{s}\right)^{N},\end{equation}
where $Z_s$ is the partition function of the subsystem. In the quasi-hermitian picture, the subsystem is described by the following Hamiltonian
 \begin{equation} \tilde{H}=\left(\begin{array}{cc}a&\epsilon\\-\epsilon&b
 \end{array}\right),\ \ a,b,\epsilon\in\mathbb{R}.
 \end{equation}
It is pseudo-hermitian with respect to $\eta$
\begin{equation} H^{\dagger}=\eta H\eta^{-1},\ \ \eta=\left(\begin{array}{cc}1&0\\0&-1
\end{array}
\right)\end{equation}
and keeps real spectrum as long as $|\epsilon|<(a-b)/2$. In this case, the mapping $\cal B$ can be constructed and reads
 \begin{equation}
 {\cal B}=\left(\begin{array}{cc}
          \frac{\epsilon}{d}& \frac{2\epsilon^2}{4\epsilon^2+(a-b)(b-a+d)}\\
	  \frac{-\epsilon}{d} & \frac{-4\epsilon^2}{-4\epsilon^2+(a-b+d)^2}	
         \end{array}\right),\ \ d=\sqrt{(a-b)^2-4\epsilon^2}. \nonumber
 \end{equation}
Let us further note that the energy levels of the system may depend on the external parameters, e.g. on the volume $v$ in which the particle is confined.

We want to find $Z_s$. It would be possible to compute the partition function exactly. However, let us pretend that the spectrum of the operator is hard to be found.
We shall solve the equation (\ref{propagator}).
Separating the Hamiltonian into its hermitian part and non-hermitian perturbation term
 \begin{equation} H=H_0+\epsilon V,\end{equation}
we introduce the following transformation
 \begin{equation} U_0(t)=\exp\left(-iH_0t\right)\end{equation}
and define a new operator
 \begin{equation} U_I(t)=U^{\dagger}_0(t)U(t)U_0(0).\end{equation}
In the interaction representation, the evolution equation (\ref{propagator}) of the system  is given by the following  \
 \begin{equation} i\frac{\partial U_I}{\partial t}=V_I(t)U_I(t),\ \ U_I(0)=1\end{equation}
where $V_I(t)=U^{\dagger}_0(t)VU_0(t)$. Integrating both parts of the equation, we can find the solution iteratively, 
% \be U_I(t)=1+\sum_{n=1}^{\infty}(-i)^n\int_{0}^tdt_1\int_{0}^{t_1}...\int_{0}^{t_{n-1}}dt_n V_I(t_1)..V_I(t_n)\end{equation}
 \begin{equation} U_I(t)=1+(-i)\epsilon\int_{0}^tdt_1V_I(t_1)-\epsilon^2\int_{0}^tdt_1\int_{0}^{t_1}dt_2 V_I(t_1)V_I(t_2)+O(\epsilon^3).\label{sol}\end{equation}
Neglecting the higher order terms, the explicit result is
 \begin{equation} U_I(t)\sim\left(\begin{array}{cc} 1+\frac{\epsilon^2}{a-b}\left(it+\frac{1-e^{i(a-b)t}}{a-b}\right)&\epsilon\frac{1-e^{i(a-b)t}}{a-b}\\
 ×\epsilon\frac{1-e^{-i(a-b)t}}{a-b}& 1-\frac{\epsilon^2}{a-b}\left(it-\frac{1-e^{-i(a-b)t}}{a-b}\right)
 \end{array}
\right).\end{equation}
The approximative value of the partition function $Z$ is
 \begin{equation} Z_s(\beta)=Tr\left(U_0(t)U_I(t)\right)|_{t\rightarrow-i\beta}\nonumber\end{equation}
\begin{equation}\sim e^{-a\beta}+e^{-b\beta}+\frac{\epsilon^2\beta}{a-b}\left(e^{-a\beta}-e^{-b\beta}\right)\label{pertZ}\end{equation}

%As we are interested in diagonal elements of the operator (), only the second correction term in (\ref{sol}) is relevant. After the integration and backward transformation and substitution and taking the trace, we get\footnote{This approximation will be sufficient as long as the parameter $\epsilon^2\beta$ is small enough. This suggests that the approximation could have problems for small temperatures, i.e. in the limit $\beta\rightarrow\infty$.}

\noindent
It would be in order to discuss the range of parameters (inverse temperature\footnote{We use atomic units, i.e. $k=1$.} $\beta=1/T$, coupling constant $\epsilon$) for which the approximation (\ref{pertZ}) is relevant. Instead, let us compare (\ref{pertZ})  with exactly computed partition function $Z^{ex}_s$ in Fig.1. 
\newsavebox{\figjedna}
    \savebox{\figjedna}{
    \rotatebox{0}{\scalebox{1}{
    \includegraphics{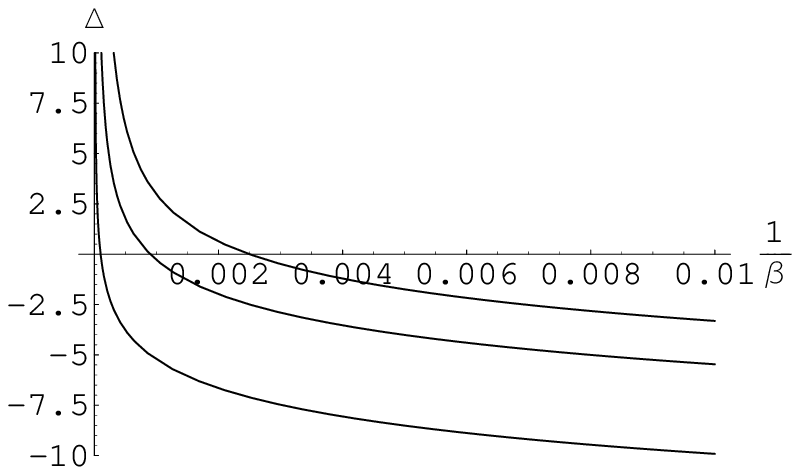}
    }}
    }
\begin{figure}
\begin{center}
\usebox{\figjedna}
\caption{Graph of function $\Delta=\text{Log}\left|\frac{Z_s-Z^{ex}_s}{Z^{ex}_s}\right|$ for $e=0.03,\ 0.02,\ 0.01$ where $ Z^{ex}_s=\text{Exp}\left(-\beta E_1\right)+\text{Exp}\left(-\beta E_2\right),\ \ E_j=\frac{1}{2}\left(a+b+(-1)^j\sqrt{(a-b)^2-4\epsilon^2}\right) \label{exactZ}.$ The approximation improves its accuracy with decreasing $e$. We set $a=1,\ b=2$.}
\end{center}
\end{figure}

Now, we can compute thermodynamic quantities. The free energy of the system is 

 \begin{equation} F=-N\frac{1}{\beta} \text{Log} Z_s=-N\frac{1}{\beta}\text{Log}\left(e^{-\beta a}+e^{-\beta b}+\epsilon^2\beta\frac{e^{-\beta a}-e^{-\beta b}}{a-b}\right)\end{equation}
\noindent
and enables a direct computation of the entropy
 \begin{equation} S=-\left(\frac{\partial F}{\partial \beta^{-1}}\right)_{v}=N\text{Log}\left[e^{-a \beta }+e^{-b \beta }+\frac{e^2 \beta\left(e^{-a \beta }-e^{-b \beta }\right)  }{a-b}\right]\nonumber\end{equation}
\begin{equation}
+N\beta\frac{  e^{a \beta } \left[\epsilon^2+b (a-b-\epsilon^2 \beta )\right]+e^{b \beta } \left[-\epsilon^2+a (a-b+\epsilon^2 \beta )\right]}{e^{a \beta } (a-b-\epsilon^2 \beta )+e^{b \beta} (a-b+\epsilon^2 \beta )} \end{equation}
\noindent
The internal energy of the system is

 \begin{equation} U=\frac{\partial\left(\beta F\right) }{\partial \beta}=N\frac{e^{\beta b}\left[a\left(a-b+\epsilon^2\beta\right)-\epsilon^2\right]+e^{\beta a}\left[b\left(a-b-\epsilon^2\beta\right)+\epsilon^2\right]}{e^{\beta a}\left(a-b-\epsilon^2\beta\right)+e^{\beta b}\left(a-b+\epsilon^2\beta\right)}.\end{equation}
The specific heat per particle $c_v$ describes how the temperature changes when the heat is absorbed and the volume $v$ of the system remains unchanged. For our system, it reads explicitly
 \begin{equation} c_v=\frac{1}{N}\left(\frac{\partial U}{\partial \beta^{-1}}\right)_{v}=\beta^2\frac{e^{\beta(a+b)}\left[(a-b)^2\left[(a-b)^2-4\epsilon^2-\epsilon^4\beta^2\right]+2\epsilon^4\right]-\epsilon^4\left(e^{2\beta a}+e^{2\beta b}\right) }{\left[e^{\beta a}\left(a-b-\epsilon^2\beta\right)+e^{\beta b}\left(a-b+\epsilon^2\beta\right)\right]^2},\label{cv}\end{equation}
see Fig 2.

The pressure $P$ of the system is defined as a derivative of free energy with respect to the volume $v$
 \begin{equation}
  P=-\left(\frac{\partial F}{\partial v}\right)_{1/\beta}\nonumber
   \end{equation}
\begin{equation}
P= 
 2N\frac{e^{\beta n/v^2}
	\left(
		m\left(
			v^2(n-m)-\beta\varepsilon^2
		 \right)
	+v^2\varepsilon^2
	\right)
	+e^{\beta m/v^2}
	\left(
		n\right(
			v^2(n-m)+\beta\varepsilon^2
		 \left)
	-v^2\varepsilon^2
	\right)}
 {  v^3\left[
	e^{\beta n/v^2}
	\left(
		v^2(n-m)-\beta\varepsilon^2
	\right)
	+e^{\beta m/v^2}
	\left(
		v^2(n-m)+\beta\varepsilon^2
	\right)
	\right]},\ \ \label{stavrce}\end{equation}
We assumed dependence of the energy levels in the following form
 $$a=\frac{n}{v^2}, \ \ b=\frac{m}{v^2}, \ \ \epsilon=\frac{\varepsilon}{v^2}. $$
\noindent
The equation (\ref{stavrce}) relates pressure, volume and temperature of the system and represents the equation of state for our system. 

\newsavebox{\figdva}
    \savebox{\figdva}{
    \rotatebox{0}{\scalebox{1}{
    \includegraphics{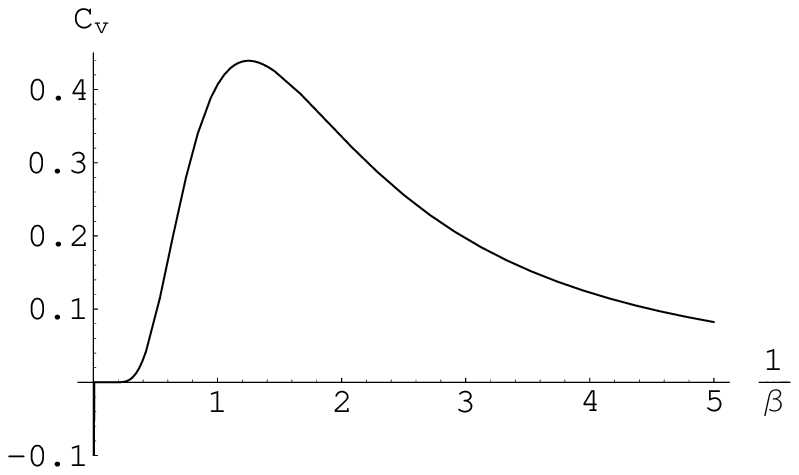}
    }}
    }
%\newsavebox{\figtri}
%    \savebox{\figtri}{
%    \rotatebox{0}{\scalebox{1}{
%    \includegraphics{formulky_gr3.eps}
%    }}
%    }
\begin{figure}
\begin{center}
\usebox{\figdva}
\caption{The specific heat (\ref{cv}) in dependence on the temperature $T=1/\beta$. We set $a=1,\ \ b=4,\ \ \epsilon=0.01$}
\end{center}
\end{figure}
%\begin{figure}
%\begin{center}
%\usebox{\figtri}
%\caption{The pressure (\ref{stavrce}) in dependence on the temperature $T=1/\beta$. We set $n=1,\ \ m=4,\ \ \varepsilon=0.01,\ \ v=2$}
%\end{center}
%\end{figure}

%In the end of the section, let us make a few remarks. The preceding analysis was based of the approximative value of partition function (\ref{pertZ}). In order to keep accuracy of the approximation, we considered weak non-hermitian interaction only, i.e. $\epsilon\sim 0$. Now, we intend to examine briefly the regimes of  stronger non-hermitian field. To do so, let us turn the attention to the exact partition function 
%  \be Z^{ex}_s=\text{Exp}\left(-\beta E_1\right)+\text{Exp}\left(-\beta E_2\right),\ \ E_j=\frac{1}{2}\left(a+b+(-1)^j\sqrt{(a-b)^2-4\epsilon^2}\right) \label{exactZ}.\ee
%We may observe a qualitative change in the behavior of the function as the coupling strength exceeds a critical value, i.e. $\epsilon>\frac{1}{2}(a-b)$. Due to the complexification of energies, it exhibits an oscillating behavior 
% \be Z^{ex}_s=2e^{-\beta \text{Re} E_1}\cos \left({\beta}\text{Im} E_1\right).\label{compZ}\ee
%\noindent
%Partition function of any physical system in equilibrium can not depend on the temperature in this way.  In the super-critical regime\footnote{Spontaneous breakdown of $PT$-symmetry in other words}, either the thermodynamics of pseudo-hermitian operators can not be build on (\ref{bloch}) and (\ref{Z}) or the Hamiltonians  can not describe physical systems in equilibrium.

\section{Discussion}
An ambition of pseudo-hermitian quantum mechanics is to serve as a consistent framework for description of physical processes, comparable with or exceeding the standard theory in predictive power. Currently, the contribution of pseudo(quasi)-hermitian operators is seen in the alternative representation of hermitian systems. Particularly, complicated hermitian Hamiltonians can be defined implicitly with help of quasi-hermitian operators which are easier to deal with \cite{faria&fring,mostafazadehdelta}. 

The construction of a mapping which relates these two operators is a crucial task of the theory and attracts a lot of attention \cite{geyer&znojil,jones&mateo,mostafazadehh,bender&brody,krejcirik&bila}. It is essential for proper probabilistic interpretation of states of non-hermitian Hamiltonian and for construction of additional operators representing physical observables.  

In the article, we considered statistical properties of an ensemble of pseudo-hermitian systems in equilibrium. Particularly, we were interested in thermodynamics of the system. It was the main observation that the explicit knowledge of the positive-definite metric operator $\Theta$ is not essential for derivation of the physical quantities. This opens a way for direct derivation of genuine physical properties of various pseudo-hermitian ($PT$-symmetric) models without tedious computation of the metric operator $\Theta$ (or, alternatively, $Q$ or $C$ operator, see \cite{Q,Cop}. Moreover, various techniques of standard quantum theory may be used without any modification. In this context, let us mention path integrals \cite{kleinert} or operator method \cite{ivanov} which might be particularly suitable for computation of partition functions of $PT$-symmetric Hamiltonians with polynomial  potentials, e.g. (\ref{ix^3}).

\section*{Acknowledgments}

The work was supported by the project No. LC06002 and by GACR grant Nr.202/07/1307.

\end{document}